\documentclass[lettersize,journal]{IEEEtran}

\usepackage[left=1.62cm,right=1.62cm,top=1.85cm,bottom=2.6cm]{geometry}

\usepackage{graphics,colortbl,multicol,lipsum,xcolor,graphicx,color}
\usepackage[cmex10]{amsmath}
\usepackage{amsthm,mathrsfs,mathtools,amsbsy,amsfonts,amssymb}
\usepackage{setspace,float,pstool,lettrine,newclude}
\usepackage[normalem]{ulem}
\usepackage{latexsym,algpseudocode,gensymb,balance,multirow,bm,wasysym}
\usepackage{cite}
\usepackage[linesnumbered,ruled,vlined]{algorithm2e}
\usepackage{bbm}
\usepackage{subfigure}
\SetKwInput{KwInput}{Input}  
\SetKwInput{KwOutput}{Output}
\SetKw{KwBy}{by}
\makeatletter
\newcommand{\eqnum}{\refstepcounter{equation}\textup{\tagform@{\theequation}}}
\newcommand{\nosemic}{\renewcommand{\@endalgocfline}{\relax}}% Drop semi-colon ;
\newcommand{\dosemic}{\renewcommand{\@endalgocfline}{\algocf@endline}}% Reinstate semi-colon ;
\let\oldnl\nl% Store \nl in \oldnl
\newcommand{\nonl}{\renewcommand{\nl}{\let\nl\oldnl}}% Remove line number for one line
\makeatother
\ifCLASSOPTIONcompsoc
\usepackage[caption=false,font=normalsize,labelfon
t=sf,textfont=sf]{subfig}
\else
\usepackage[caption=false,font=footnotesize]{subfig}
\fi
\DeclarePairedDelimiterX\MeijerM[3]{\lparen\!}{\rparen}%
{\,#3\delimsize\vert\begin{smallmatrix}#1 \\ #2\end{smallmatrix}}

\newcommand\MeijerG[8][]{%
  G^{\,#2,#3}_{#4,#5}\MeijerM[#1]{#6}{#7}{#8}}

\WithSuffix\newcommand\MeijerG*[7]{%  
  G^{\,#1,#2}_{#3,#4}\MeijerM*{#5}{#6}{#7}}
\allowdisplaybreaks
\newtheorem{remark}{Remark}

\newtheorem{proposition}{Proposition}

\graphicspath{{figures/}}
\allowdisplaybreaks

\newcommand{\RNum}[1]{\uppercase\expandafter{\romannumeral #1\relax}}

\usepackage{mathtools}

\newcommand{\probP}{\mathrm{Pr}}

\usepackage{accents}

\begin{document}

\title{Goal-Oriented Multiple Access Connectivity\\ for Networked Intelligent Systems}
\author{Pouya Agheli, 
\IEEEmembership{Graduate Student Member, IEEE}, Nikolaos Pappas, 
\IEEEmembership{Senior Member, IEEE},\\ and Marios Kountouris, \IEEEmembership{Fellow, IEEE}.
\thanks{P. Agheli and M. Kountouris are with the Communication Systems Department, EURECOM, Sophia-Antipolis, France, email: \{\texttt{pouya.agheli, kountour\}@eurecom.fr}. M. Kountouris is also with the Department of Computer Science and Artificial Intelligence, University of Granada, Spain. N. Pappas is with the Department of Computer and Information Science, Linköping University, Sweden, email: \texttt{nikolaos.pappas@liu.se}. The work of P. Agheli and M. Kountouris has received funding from the European Research Council (ERC) under the European Union’s Horizon 2020 research and innovation programme (Grant agreement No. 101003431). The work of N. Pappas has been supported by the Swedish Research Council (VR), ELLIIT, and the European Union (ETHER, 101096526).}}
\maketitle

\begin{abstract}
We design a self-decision goal-oriented multiple access scheme, where sensing agents observe a common event and individually decide to communicate the event's attributes as updates to the monitoring agents, to satisfy a certain goal. Decisions are based on the usefulness of updates, generated under uniform, change- and semantics-aware acquisition, as well as statistics and updates of other agents. We obtain optimal activation probabilities and threshold criteria for decision-making under all schemes, maximizing a grade of effectiveness metric. Alongside studying the effect of different parameters on effectiveness, our simulation results show that the self-decision scheme may attain at least $92\%$ of optimal performance.
\end{abstract} 

\IEEEpeerreviewmaketitle

\section{Introduction}
\lettrine{G}{oal-oriented} semantic communication holds great promise for realizing resource-efficient and scalable networked intelligent systems \cite{kountouris2021semantics}, where sensing agents observe events and convey the events' attributes to monitoring agents to achieve a certain \emph{goal}. Therein, update acquisition, data transmission, and information usage become effective in satisfying the goal when only useful or significant updates are considered in the communication lifecycle. The challenge is exacerbated when the medium is shared, and agents have to communicate over multiple access channels. Conventional medium access protocols, such as typed-based ALOHA, grant-based, and grant-free access, are primarily goal- and effectiveness-agnostic, sending all frames regardless of their goal-dependent usefulness and impact at the endpoint. Despite some prior work \cite{mergen2006type, liu2018sparse, wu2020massive, munari2021modern, choi2021grant, daei2023blind}, the problem of goal-oriented multiple access and update provision remains mostly unexplored.

In this letter, we propose a \emph{self-decision goal-oriented} multiple access scheme, in which each sensing agent decides to \emph{speak up} or \emph{remain silent} based on the usefulness of its generated updates and the statistics of the other agents and their generated updates for achieving a specific goal. The design involves computing the optimal activation probabilities of the agents and a decision-making criterion, which maximizes a \emph{grade of effectiveness} metric that integrates the discrepancy error between actual and reconstructed events, the consumed resources, and the usefulness of perceived updates. Our results indicate that the proposed self-decision scheme is very close to the optimal solution.

\section{System Model}
\begin{figure}[t!]
\centering
\pstool[scale=0.35]{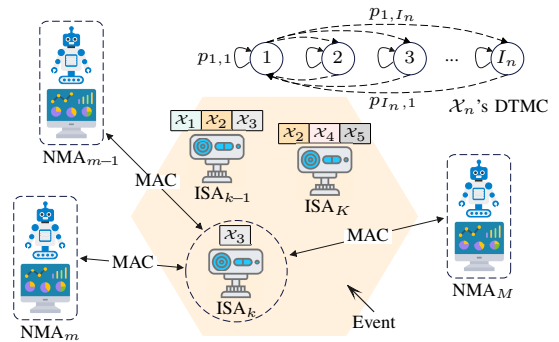}{
\psfrag{ISA1}{\hspace{-0.15cm}\scriptsize $\text{ISA}_{k\!-\!1}$}
\psfrag{ISA2}{\hspace{-0.1cm}\scriptsize $\text{ISA}_k$}
\psfrag{ISAK}{\hspace{-0.1cm}\scriptsize $\text{ISA}_K$}
\psfrag{NMA1}{\hspace{-0.18cm}\scriptsize $\text{NMA}_{m\!-\!1}$}
\psfrag{NMA2}{\hspace{-0.13cm}\scriptsize $\text{NMA}_m$}
\psfrag{NMAM}{\hspace{-0.13cm}\scriptsize $\text{NMA}_M$}
\psfrag{Pheno}{\hspace{-0.75cm}\scriptsize Event}
\psfrag{SourceDTMC}{\hspace{0.3cm}\scriptsize $\mathcal{X}_n$'s DTMC}
\psfrag{AL1}{\hspace{-0.05cm}\tiny $\mathcal{X}_1$}
\psfrag{AL2}{\hspace{-0.05cm}\tiny $\mathcal{X}_2$}
\psfrag{AL3}{\hspace{-0.05cm}\tiny $\mathcal{X}_3$}
\psfrag{AL4}{\hspace{-0.05cm}\tiny $\mathcal{X}_4$}
\psfrag{AL5}{\hspace{-0.05cm}\tiny $\mathcal{X}_5$}
\psfrag{1}{\hspace{-0.16cm} \scriptsize $1$}
\psfrag{2}{\hspace{-0.16cm} \scriptsize $2$}
\psfrag{3}{\hspace{-0.16cm} \scriptsize $3$}
\psfrag{L}{\hspace{-0.24cm} \scriptsize $I_n$}
\psfrag{p00}{\hspace{-0.32cm} \scriptsize $p_{1,1}$}
\psfrag{p01}{\hspace{-0.25cm} \scriptsize $p_{1,I_n}$}
\psfrag{p10}{\hspace{-0.25cm} \scriptsize $p_{I_n,1}$}
\psfrag{MAClinkx}{\hspace{-0.09cm} \scriptsize MAC}
\psfrag{...}{\hspace{-0.26cm} \scriptsize ...}
}
\caption{Goal-oriented medium access in a networked intelligent system.}
\label{Fig:Figure0}
\end{figure}

We consider $K$ intelligent sensing agents (ISAs) that observe a common event (information source) and convey their noisy observations in the form of updates to $M$ networked monitoring agents (NMAs) over a multiple access wireless channel prone to errors and universal frequency reuse, c.f., Fig.~\ref{Fig:Figure0}. Communication occurs over independent time-slotted service intervals. Within each interval, all NMAs \emph{query} for new observations from the ISAs in the downlink, and the ISAs respond to those queries in the uplink, providing updates within the same service interval. The evolution of the source $X_t$ at the $t$-th service interval, with $t\in\mathbb{N}$, belongs to a finite state space. An event at each state is described via a set of \emph{information attributes}, defined by $\mathcal{A}=\{\mathcal{X}_n~|~n=1, 2, \dots, N\}$. Here, $\mathcal{X}_n \in \{x_i^{(n)}~|~i=1, 2, \dots, I_n\}$ is a level-$n$ attribute, which is modeled by a discrete-time Markov chain (DTMC) with $I_n\geq2$ states. In the DTMC modeling the $n$-th attribute, we assume $p_{i,i} = p_n^\prime$, and $p_{i,i^\prime} = p_n$, $\forall i \neq i^\prime$, for $i,i^\prime = 1, 2, \dots, I_n$, where $p_n^\prime + (I_n\!-\!1)p_n = 1$.

At the $t$-th interval, the NMAs query for a group of attributes $\widehat{\mathcal{A}}_t \subseteq \mathcal{A}$ represented by the sequence $\langle \widehat{\mathcal{A}}_t \rangle$, where $|\widehat{\mathcal{A}}_t| \leq N$. Either queried attribute is independently communicated in one slot $j$, $j=1,2, \dots, |\widehat{\mathcal{A}}_t|$, with the same order as in $\langle \widehat{\mathcal{A}}_t \rangle$. The levels of queried attributes are determined in line with the goal to which the NMAs subscribe. The $k$-th ISA, where $k=1,2, \dots, K$, can observe only a subset of attributes $\mathcal{A}_k \subseteq \mathcal{A}$ and generate updates from its noisy observations via an encoder, yielding a codebook shared with the other agents. Moreover, $\mathcal{K}_n = \langle k: \mathcal{X}_n \in \mathcal{A}_k \rangle$ is the sequence of ISAs that can observe the $n$-th attribute. To estimate the event, the NMAs need to jointly decode all queried attributes and construct $\widehat{X}_t$ at the $t$-th service interval. An attribute is correctly perceived at that interval if at least $M_t$ NMAs successfully decode it, where $1\leq M_t \leq M$.

\subsection{Goal-Oriented Update Provision}\label{Section2:2a}
At the time slot order of the $n$-th queried attribute, the $k$-th ISA, $\forall k \in \mathcal{K}_n$, generates an update from its observation with probability $\beta_{k,n}$, as obtained in Section~\ref{Section4:4d}, under the following acquisition schemes \cite{pappas2021goal}:
\begin{enumerate}
    \item \textit{Uniform:} An update is generated periodically at each time slot, regardless of the prior state of the source and the latest perceived update at the NMAs.
    \item \textit{Change-aware:} An update for an attribute is generated if the state of that attribute changes at a time slot, regardless of the latest perceived update.
    \item \textit{Semantics-aware:} Update generation for an attribute is triggered once there is a discrepancy between the current actual state of the attribute and that of the latest decoded one of the same level at the NMAs. In this case, the ISAs should maintain a record of the last decoded attributes.
\end{enumerate}

The ISA follows a self-decision policy, thanks to which it individually decides whether to transmit that update relevant to the $n$-th attribute with probability $\alpha_{k,n}$. Each update is assigned a meta value according to its usefulness in satisfying the subscribed goal \cite{agheli2023semantic}. We consider $v_{k, i}^{(n)}$, $\forall k,i,n$, the meta value for the realization $x_i^{(n)}$ at the $k$-th ISA, and assume it is assigned based on a knowledge fusion model, as in \cite[Section~III-A]{agheli2023semantic}. In this work, the self-decision scheme takes the form of a threshold-based decision policy. Thus, the $k$-th ISA speaks up at the $j$-th slot if $v_{k,j}^{(n)} > v_{{\rm th}, k}^{(n)}$; otherwise, it remains silent. The threshold $v_{{\rm th}, k}^{(n)}$, shows a variable \emph{decision-making criterion} for the $k$-th ISA observing the $n$-th attribute.

\subsection{Transmission Success Probability}
As observations are noisy, updates generated from the same event via different ISAs can be different, which leads to signal \emph{contamination} over the wireless network. We consider $h_{k, m} = \bar{h}_{k, m} d_{k, m}^{-\frac{a}{2}}$ the channel between the $k$-th ISA and the $m$-th NMA, represented by an independent and identically distributed (i.i.d.) random variable (r.v.) where $\bar{h}_{k, m}$ is the small-scale fading with its power having a cumulative distribution function (CDF) of $F_{h^2}(|\bar{h}_{k, m}|^2) = 1 - \operatorname{exp}(-|\bar{h}_{k, m}|^2)$, $d_{k, m}$ denoting the i.i.d. distance r.v.s, and $2\leq a < 7$ being the path-loss exponent. Let us consider $\gamma_{\rm th}$ the signal-to-noise ratio (SNR) sensitivity. The probability of successful transmission of the update from the $k$-th ISA and received at the $m$-th NMA is $\gamma_{k, m, n}^{(\mathcal{C}, \mathcal{I})} = \probP\big(\operatorname{SNR}_{k, m,n}^{(\mathcal{C}, \mathcal{I})} > \gamma_{\rm th}\big)$. The channel drop-off occurs when $\operatorname{SNR}_{k, m,n}^{(\mathcal{C}, \mathcal{I})} \leq \gamma_{\rm th}$, with the probability of $1-\gamma_{k, m, n}^{(\mathcal{C}, \mathcal{I})}$. Here, $\mathcal{C}$ shows the set of collaborating ISAs, and $\mathcal{I}$ is that of interfering ones. After some algebraic manipulations, we reach
\begin{align}\label{eq:eq4}
    \gamma_{k, m, n}^{(\mathcal{C}, \mathcal{I})} &= \sum_{k^{\prime} \in \mathcal{C}_k} \sum_{k^{\prime\prime} \in \mathcal{I}} \Bigg[\dfrac{\prod_{k^{\prime}_0 \in \mathcal{C}_k} \Lambda_{k^{\prime}_0, m}^{(n)} \prod_{k^{\prime\prime}_0 \in \mathcal{I}} \Omega_{k^{\prime\prime}_0, m}^{(n)} }{\big(\Lambda_{k^{\prime}, m}^{(n)} \!+\! \Omega_{k^{\prime\prime}, m}^{(n)}\big)\Lambda_{k^{\prime}, m}^{(n)} \Psi_{k,k^\prime, k^{\prime\prime}, m}^{(n)}} \nonumber \\ 
    &~~~~~~ \times \operatorname{exp}\!\left(-\Lambda_{k^{\prime}, m}^{(n)} \gamma_{\rm th} \sigma_m^2 \right)\! \Bigg] 
\end{align}
where $\mathcal{C}_k = \mathcal{C}\cup\{k\}$, $\Lambda_{k^{\prime}, m}^{(n)} = \frac{d_{k^\prime, m}^{a}}{\rho_{k^\prime, j}^{(n)}}$, $\Omega_{k^{\prime\prime}, m}^{(n)} = \frac{d_{k^{\prime\prime}, m}^{a}}{\gamma_{\rm th}\rho_{k^{\prime\prime}, j}^{(n)}}$,

\begin{align*}
     \Psi_{k, k^{\prime}, k^{\prime\prime}, m}^{(n)}  &=  \prod_{i \in \mathcal{C}_k \setminus\{k^\prime\}} \! \!\!\left(\Lambda_{i, m}^{(n)} \!-\! \Lambda_{k^{\prime}, m}^{(n)}\right)\! \\ \nonumber 
     &~~~\times \!\!\prod_{j \in \mathcal{I}\setminus\{k^{\prime\prime}\}}\! \!\!\left(\Omega_{j, m}^{(n)} \!-\! \Omega_{k^{\prime\prime}, m}^{(n)}\right)\!,
\end{align*}
$\sigma_m^2$ is the power of the complex additive white Gaussian noise (AWGN) at the $m$-th NMA, and $\rho_{k, j}^{(n)}$ is the transmission power of the $k$-th ISA for the $n$-th attribute at the $j$-th time slot. 

\section{Performance Metrics}\label{section3}
We propose a grade of effectiveness (GoE) metric to evaluate the effectiveness of communicated updates in the form of ${\rm GoE} = \varphi\left(\mathbf{W}^T; g_1(f_1), \dots, g_R(f_R)\right)$, which has $R \in \mathbb{Z}^+$ effectiveness features, with $f_r \in \mathbb{R}_0^+$ for $r=1, 2, \dots, R$ denoting the $r$-th one. Herein, $\mathbf{W} = [w_1, \dots, w_R]^T \in \mathbb{R}^{R}$, $0\leq w_r\leq 1$, indicates a weight factor, where $\mathbf{W}^T \mathbf{1}_{R\times1} = 1$. Also, $g_r : \mathbb{R}_0^+ \to \mathbb{R}$ is a positive non-decreasing differentiable function, and $\varphi: \mathbb{R}^{2R} \to \mathbb{R}$ is a context-dependent utility function \cite{kountouris2021semantics}. In this letter, we focus on three effectiveness features to determine the GoE: \emph{Effective discrepancy error} (EDE), \emph{effective resource consumption} (ERC), and \emph{effective utility of updates} (EUU). 

\subsection{Effective Discrepancy Error (EDE)}
Update discrepancy error $E_t$ occurs at the $t$-th service interval if the actual event $X_t$ is not equal to the reconstructed $\widehat{X}_t $, i.e., $E_t= \mathbbm{1} \{ X_t \neq \widehat{X}_t\}$. The EDE depicts the feature that the effect of losing information is scaled by its usefulness in satisfying the goal. Given this, we model the EDE as follows
\begin{align}\label{eq:eq2}
    {\rm EDE}(E_t, \widehat{\mathcal{A}}_t; \pmb{\alpha}, \mathbf{v}) =  \mathcal{F}_1 \sum_{n: \mathcal{X}_n \in \widehat{\mathcal{A}}_t}P_{{\rm e}, n} 
\end{align}
\vspace{-0.2cm}
\begin{align*}
    \text{where~} \mathcal{F}_1 &= \sum_{j=1}^{{\mid\widehat{\mathcal{A}}_t\mid}} \left[\, \sum_{k\in\mathcal{K}_n} \alpha_{k,n} v_{k, j}^{(n)}\right]_{n: \mathcal{X}_n= \mathcal{X}_j},
\end{align*}
$\pmb{\alpha} = [\alpha_{k,n}]_{K\times N}$, $\mathbf{v} = [v_{k,j}^{(n)}]_{K\times \mid\widehat{\mathcal{A}}_t\mid \times N}$, $\mathcal{X}_j \in \widehat{\mathcal{A}}_t$ indicates the attribute queried at the $j$-th time slot. We also have
\begin{align}\label{eq:eq3}
    P_{{\rm e}, n} = \underset{T \to \infty}{\lim} \frac{1}{T} \sum_{t=1}^{T} \bigg[\probP\!\left({E}_{t,j}^{(n)}=1\right)\!\bigg]_{j:\mathcal{X}_j \in \widehat{\mathcal{A}}_t},
\end{align}
which is the expected discrepancy error probability for the $n$-th attribute and is derived according to Proposition~\ref{prob1}. In \eqref{eq:eq3}, we define ${E}_{t,j}^{(n)}=\mathbbm{1} \{x_j^{(n)} \neq \widehat{x}_j^{(n)}~|~t\}$, and $\widehat{x}_j^{(n)}$ shows the decoded realization via the NMAs for the $n$-th attribute at the $j$-th slot.

\begin{proposition}\label{prob1} The expected discrepancy error probability $P_{{\rm e}, n}$ for the $n$-th attribute is derived as
\begin{align}\label{prob1:eq1}
    P_{{\rm e}, n} =\dfrac{\mathcal{E}_n (I_n\!-\!1)p_n}{1 + \mathcal{E}_n\!\left[ 2(I_n\!-\!1)p_n - \frac{I_n-2}{I_n-1}p_n -1 \right]}
\end{align}
where $\mathcal{E}_n=\frac{1}{2^{|\mathcal{K}_n|}} \sum_{\ell=1}^{2^{|\mathcal{K}_n|}} \mathcal{E}_{n,\ell}$ denotes the probability that the $n$-th attribute is not successfully delivered to $M_t$ NMAs, where
\begin{align}\label{prob1:eq2}
    &\mathcal{E}_{n,\ell} =  \prod_{k_0 \in \mathcal{R}^\prime_{\ell}}(1 \!-\! \alpha_{k_0, n}) \prod_{k_1 \in \bar{\mathcal{R}}_{\ell}} \alpha_{k_1, n}  \nonumber \\ 
    &~~\times\! \sum_{\tau=1}^{2^{|\mathcal{R}_{\ell}|}} \!\Bigg[
    \prod_{k_2 \in \mathcal{T}^\prime_\tau}(1 \!-\! q_{k_2, n}) \prod_{k_3 \in \mathcal{T}_{\tau}} q_{k_3, n}
     \prod_{m \in \mathcal{M}_{k_3}} \!\!\left(1\!-\!\gamma_{k_3, m, n}^{(\mathcal{T}_\tau, \mathcal{T}_\tau^\prime)}\right)\!\Bigg]
\end{align}
where $q_{k,n}$ is the probability that the $k$-th ISA's observation is correct. Moreover, $\mathcal{R}_\ell$, $\mathcal{R}^\prime_\ell$, $\mathcal{T}_\tau$, and $\mathcal{T}^\prime_\tau$ are the sets of ISAs that are active, inactive, active with correct observations, and active with incorrect observations, respectively. Also, $\mathcal{M}_{k_3}$ shows the subset of $M_t+1$ NMAs being farthest from the $k_3$-th ISA.
\end{proposition}
\begin{IEEEproof} We use the same approach as in \cite[Section~3.2]{pappas2021goal} to derive \eqref{prob1:eq1}. Considering its definition, the discrepancy error $ E_{t, j}^{(n)} $, $\forall t,j,n$, follows a DTMC with two possible states ${\{0,1\}}$ and transition probabilities $\pi^{(n)}_{a,b} = \probP\big(E_{t+1, j^\prime}^{(n)} = b~|~E_{t, j }^{(n)} = a\big)$, $\forall a,b \in \{0,1\}$, where $\mathcal{X}_j=\mathcal{X}_{j^\prime}=\mathcal{X}_n$. Therefore, we can obtain the steady-state error probability $P_{{\rm e}, n}$, as follows
\begin{align}\label{prob1:eq3}
    P_{{\rm e}, n} = \dfrac{\pi^{(n)}_{0,1}}{1 + \pi^{(n)}_{0,1} - \pi^{(n)}_{1,1}}.
\end{align}

After computing $\pi^{(n)}_{0,1}$ and $\pi^{(n)}_{1,1}$, we insert them into \eqref{prob1:eq3} and obtain \eqref{prob1:eq1} after some algebraic manipulations.
\end{IEEEproof} 

\vspace{-0.045cm}
\subsection{Effective Resource Consumption (ERC)}
The ERC feature allows us to model the resource consumption for communicating updates as a function of its usefulness. Having time-slotted channels with shared frequency band, \emph{power} is the main candidate resource to adapt. Therefore, we define the transmission power of the $k$-th ISA at the $j$-th time slot as $\rho_{k, j}^{(n)} = f_\rho(v_{k, i}^{(n)})$, in the form of a non-decreasing function $f_\rho : \mathbb{R}^+_0 \to \mathbb{R}^+_0$ of the meta value assigned to $x_i^{(n)}$ at the $j$-th slot. In this regard, we can write
\begin{align}\label{eq:eq12}
    &{\rm ERC}(\widehat{\mathcal{A}}_t; \pmb{\alpha}, \mathbf{v}) = \mathcal{F}_2 \prod_{n: \mathcal{X}_n \in \widehat{\mathcal{A}}_t}\mathcal{S}_n
\end{align}
\vspace{-0.2cm}
\begin{align*}
    \text{where~} \mathcal{F}_2 
    = \sum_{j=1}^{{\mid\widehat{\mathcal{A}}_t\mid}} \left[\sum_{k\in\mathcal{K}_n} h\!\left(\alpha_{k, n} \dfrac{f_\rho(v_{k, j}^{(n)})}{v_{k, j}^{(n)}}\right)\right]_{n :\mathcal{X}_n =\mathcal{X}_j}\!,
\end{align*}
$h : \mathbb{R}^+_0 \to \mathbb{R}^+_0$ is a non-decreasing differentiable function, and $\mathcal{S}_n = 1 - \mathcal{E}_n$ denotes the probability of correctly delivering  the $n$-th attribute to the required number of NMAs.

\vspace{-0.045cm}
\subsection{Effective Utility of Updates (EUU)}
The EUU feature shapes the utility (usefulness) of updates in satisfying the goal once the actual event they are representing is correctly reconstructed at the NMAs. Thereby, the EUU at each service interval depends on the sum value of the communicated attributes at that period, modeled as follows
\begin{align}\label{eq:eq11}
    {\rm{EUU}}(\widehat{\mathcal{A}}_t; \pmb{\alpha}, \mathbf{v}) = \mathcal{F}_1 \prod_{n: \mathcal{X}_n \in \widehat{\mathcal{A}}_t}\mathcal{S}_n.
\end{align}

\section{Multiple Access Design}\label{sec:sec4}
We aim to design a multiple access scheme via obtaining the set of optimal activation probabilities $\pmb{\alpha}^* = [\alpha_{k,n}^*]_{K\times N}$, maximizing the GoE in the system based on the goal. Next, we find the threshold criteria for decision-making and propose a self-decision scheme under all update acquisition schemes.

\subsection{Problem Formulation}\label{section4:partA}
Based on Section~\ref{section3}, we formulate the problem as follows
\begin{align}\label{eq:eq13}
    &\mathcal{P}_1 :~
    \underset{\pmb{\alpha},\mathbf{W}}{\text{max}}~
    {\rm GoE} = \varphi\left(\mathbf{W}^T; g_1(f_1), \dots, g_R(f_R)\right) \nonumber \\
    &\text{s.t.} ~~~0 \preceq \pmb{\alpha} \preceq 1,\, 0 \prec \mathbf{W} \preceq 1,\, \mathbf{W}^T \mathbf{1}_{R\times1} = 1.
\end{align}

Without loss of generality, we transform the GoE optimization in $\mathcal{P}_1$ to a problem of minimizing the weighted sum of the EDE and ERC features, named $\mathcal{G}$, subject to guaranteeing that the EUU exceeds a minimum value denoted by ${\rm{EUU}}_{\rm min}$ at each service interval. As transmissions are performed independently among the service intervals, it is sufficient to analyze the effectiveness problem for an arbitrary $t$-th period. Then, we can apply the same access design for the other periods, depending on $\widehat{\mathcal{A}}_t$, $\forall t$. Therefore, we can write
\begin{align}\label{eq:eq14}
    &\mathcal{P}_2 :~
    \underset{0 \preceq \pmb{\alpha} \preceq 1,\, 0 \prec \mathbf{W} \preceq 1}{\text{min}}
    ~\mathcal{G} = \sum_{r=1}^{2}w_rg_r(f_r) \nonumber \\
    &\text{s.t.}~\, g_3\big({\rm{EUU}}(\widehat{\mathcal{A}}_t; \pmb{\alpha}, \mathbf{v})\big) \geq {\rm{EUU}}_{\rm min},~ \mathbf{W}^T \mathbf{1}_{3\times1} = 1
\end{align}
where $f_1 = {\rm EDE}(E_t, \widehat{\mathcal{A}}_t; \pmb{\alpha}, \mathbf{v})$, and $f_2={\rm ERC}(\widehat{\mathcal{A}}_t; \pmb{\alpha}, \mathbf{v})$.

\vspace{-0.15cm}
\subsection{Problem Solution}\label{section4:partB}
Inserting \eqref{eq:eq2}, \eqref{eq:eq12}, and \eqref{eq:eq11} into \eqref{eq:eq14}, we define the Lagrange function $\mathcal{L}$ for $\mathcal{P}_2$, and we can write the Karush-Kuhn-Tucker (KKT) conditions, as follows
\begin{align}\label{eq:eq18}
    &\dfrac{\partial \mathcal{L}}{\partial \alpha_{k,n}}  = \nonumber \\
    &~~~w_1 \Bigg[ \mathcal{F}_1 \!\left(\dfrac{\partial P_{{\rm e}, n}}{\partial \alpha_{k,n}}\right)\! +\! P_{{\rm e}, n} \!\left(\dfrac{\partial \mathcal{F}_1}{\partial \alpha_{k,n}}\right)\! \Bigg]  g_1^\prime\!\!\left( \mathcal{F}_1 \sum_{n: \mathcal{X}_n \in  \widehat{\mathcal{A}}_t} P_{{\rm e}, n} \right) \nonumber \\
    &~~~+ w_2 \overline{\mathcal{S}}_n \Bigg[\mathcal{S}_n \!\left( \dfrac{\partial \mathcal{F}_2}{\partial \alpha_{k,n}} \right)\!+\! \mathcal{F}_2\!\left( \dfrac{\partial \mathcal{S}_n}{\partial \alpha_{k,n}} \right)\!\Bigg] g_2^\prime\!\left(\mathcal{F}_2 \mathcal{S}_n\overline{\mathcal{S}}_n\right) \nonumber\\
    &~~~- \eta \overline{\mathcal{S}}_n \Bigg[ \mathcal{S}_n \!\left( \dfrac{\partial \mathcal{F}_1}{\partial \alpha_{k,n}} \right)\!+\! \mathcal{F}_1 \!\left( \dfrac{\partial \mathcal{S}_n}{\partial \alpha_{k,n}} \right) \!\Bigg] g_3^\prime\!\left( \mathcal{F}_1 \mathcal{S}_n\overline{\mathcal{S}}_n  \right) \nonumber \\
    &~~~+ \lambda_{k,n} = 0,\, \forall k,n,
\end{align}
where $ \overline{\mathcal{S}}_n =\!\! \prod_{n^\prime \in \widehat{\mathcal{A}}_t \setminus {\{\mathcal{X}_n\}}}\!\mathcal{S}_{n^\prime}$.
Moreover, we have
\begin{align}\label{eq:eq19}
    \dfrac{\partial \mathcal{L}}{\partial w_1} &= g_1\!\!\left( \mathcal{F}_1 \sum_{n: \mathcal{X}_n \in  \widehat{\mathcal{A}}_t} P_{{\rm e}, n}\right)\! + \mu_0 + \mu_1 = 0, \\ \label{eq:eq20}
    \dfrac{\partial \mathcal{L}}{\partial w_2} & =  g_2\!\left(\mathcal{F}_2 \mathcal{S}_n\overline{\mathcal{S}}_n\right)\! + \mu_0 + \mu_2 = 0.
\end{align}
In \eqref{eq:eq18}--\eqref{eq:eq20}, $\eta \geq 0$, $\mu_r \geq 0$, $\forall r$, and $\lambda_{k,n} \geq 0$, $\forall n: \mathcal{X}_n \in \widehat{\mathcal{A}}_t$, $\forall k \in \mathcal{K}_n$, denote Lagrange multipliers.

The complementary slackness conditions are as below
\begin{align*}
    &\mathcal{H}_1: \eta \Big(g_3\!\left( \mathcal{F}_1 \mathcal{S}_n\overline{\mathcal{S}}_n\right) \!- {\rm{EUU}}_{\rm min}  \Big) = 0, \\
     \mathcal{H}_2 : \mu_r \!\left(w_r\!-\!1\right)\!&=0, \, r=1,2, ~\,\mathcal{H}_3:\lambda_{k,n}\!\left(\alpha_{k,n}\!-\!1\right)\!=0,\, \forall k,n .
\end{align*}
Each of $\mathcal{H}_1$ to $\mathcal{H}_3$ holds under two conditions: In $\mathcal{H}_3$, we consider (i) $\lambda_{k,n} = 0$, hence $\alpha_{k,n} < 1$; or (ii) $\alpha_{k,n} = 1$, thus $\lambda_{k,n} > 0$. Under condition (ii), the ISAs are active at all time slots, which is not optimal according to \eqref{prob1:eq1}. Thus, we reach $\lambda_{k,n}=0$, $\forall k,n$. In $\mathcal{H}_2$, we assume (i) $\mu_{r} = 0$, then $w_r < 1$; or (ii) $w_r = 1$, hence $\mu_{r} > 0$. Condition (ii) leads one of the weighting variables to become zero, which is incorrect referring to $\mathcal{P}_2$. Thus, we have $\mu_r=0$, $\forall r$. Given this, we arrive
\begin{align}\label{eq:eq21}
    \mathcal{F}_2 =  \dfrac{1}{ \mathcal{S}_n\overline{\mathcal{S}}_n} g_2^{-1}\!\!\left[g_1\!\!\left( \mathcal{F}_1 \sum_{n: \mathcal{X}_n \in  \widehat{\mathcal{A}}_t}P_{{\rm e}, n}\right)\! \right]\!.
\end{align}
Also, for $\mathcal{H}_1$, we have (i) $\eta = 0$, and $g_3\!\left( \mathcal{F}_1 \mathcal{S}_n\overline{\mathcal{S}}_n \right) > {\rm{EUU}}_{\rm min} $; or (ii) $g_3\!\left( \mathcal{F}_1 \mathcal{S}_n\overline{\mathcal{S}}_n \right) = {\rm{EUU}}_{\rm min}$, and $\eta > 0$. Applying \eqref{eq:eq21} into \eqref{eq:eq18}, condition (i) results in negative probabilities. Thus, condition (ii) is true with $g_3\!\left( \mathcal{F}_1 \mathcal{S}_n\overline{\mathcal{S}}_n \right) = {\rm{EUU}}_{\rm min}$. From \eqref{eq:eq18}, we obtain
\begin{align}\label{eq:eq23}
    P_{{\rm e}, n} = \left[ \dfrac{g_1^{-1}\!\!\left(\dfrac{\eta \mathcal{S}_n \overline{\mathcal{S}}_n {\rm{EUU}}_{\rm min}}{2w_1}  \right)}{\mathcal{F}_1 \sum_{n^\prime: \mathcal{X}_{n^\prime} \in \widehat{\mathcal{A}}_t \setminus {\{\mathcal{X}_n\}}} P_{{\rm e}, n^\prime}} \right]^{\!1/2} \!.
\end{align}

Putting together \eqref{prob1:eq1}, \eqref{prob1:eq2}, and \eqref{eq:eq23}, and after some simple numerical and algebraic manipulations based on the forms of $g_1$, $h$, and $f_\rho$, the optimal $\alpha^*_{k,n}$, $\forall k,n$, is derived. We achieve optimal $w_1^*$ and $w_2^*$ by exerting $\pmb{\alpha}^*$ into the Lagrange function and equalizing the equation $\mathcal{L}(\mathbf{W}) = \mathcal{L}(\pmb{\alpha}^*)$. In this regard, Algorithm~\ref{Alg:Alg.1} is proposed, which provides the optimal values for the optimization variables with certain stopping accuracy and convergence rate of $\mathcal{O}((N_iN_j)^{-1})$, where $N_i$ and $N_j$ are the maximum numbers of inner and outer loops, respectively.

\begin{algorithm}[t!]
\DontPrintSemicolon
    \caption{Solution for deriving $\pmb{\alpha}^*$ and $\mathbf{W}^*$ at the $t$-th service interval.} \label{Alg:Alg.1}
    \KwInput{Known parameters $M$, $L$, $p$, $\widehat{\mathcal{A}}_t$, $\mathcal{K}_n$, $v_{k,j}^{(n)}$, $\forall j, k$, $q_{k,n}$, $\gamma_{k,m,n}^{(\mathcal{C}, \mathcal{I})}$, $\forall n, k$, ${\rm{EUU}}_{\rm min}$, $\pmb{\varepsilon_1}$, and $\pmb{\varepsilon_2}$. Initial parameters $\eta^{(0)} \gets 0$, $\pmb{\alpha}^{(0)}=\mathbf{0}_{K\times|\widehat{\mathcal{A}}_t|}$, and $\mathbf{W}=[0.5, 0.5]^T$. Forms of $g_r$, $\forall r$, $h$, and $f_\rho$.
    }
    \KwOutput{Optimal parameters $\alpha^*_{k,n}$, $\forall k,n$, $w_1^*$, and $w_2^*$.}
    \textit{Iteration} $j$: \Comment{Outer loop}\\
    \lIf{conditions $\mathcal{H}_4$ to $\mathcal{H}_6$ are met.}{\textbf{goto} {\scriptsize{\textbf{3}}}.}
    \textit{Iteration} $i$: \Comment{Inner loop}\\
     Consider $n=\langle \widehat{\mathcal{A}}_t(c) \rangle_{c=1}$ and $k = \langle \mathcal{K}_n(c) \rangle_{c=1}$.\\
     Update $\overline{\mathcal{S}}_n^{(i)}$, $\mathcal{F}_1^{(i)}$, $\mathcal{F}_2^{(i)}\!(\pmb{\alpha})$, and $P^{(l,i)}_{{\rm e}, n^\prime}$, $\forall n^\prime\in\widehat{\mathcal{A}}_t \setminus \{n\}$. \\   
     Derive $\alpha_{k,n}^*$ from \eqref{prob1:eq1}, \eqref{prob1:eq2}, \eqref{eq:eq23}, and given $\alpha_{k^\prime,n}$, $\forall n,k^\prime \in \mathcal{K}_n \setminus \{k\}$; update and save $\pmb{\alpha}^{(i)}$.\\
     \textbf{goto} {\scriptsize{\textbf{4}}} and repeat for $\langle \widehat{\mathcal{A}}_t(c) \rangle_{c=2}^{|\widehat{\mathcal{A}}_t|}$ and $\langle \mathcal{K}_n(c) \rangle_{c=2}^{|\mathcal{K}_n|}$.\\
     \lIf{$\Big|\pmb{\alpha}^{(i)} \!-\! \pmb{\alpha}^{(i-1)}\Big|\!>\! \pmb{\varepsilon_1}$}{set $i\!=\!i\!+\!1$, and \textbf{goto} {\scriptsize{\textbf{3}}}.}
     Compute $w_1^{(j)}$ and $w_2^{(j)}$; save $\mathbf{W}^{(j)}$.\\
     \lIf{$\Big|\mathbf{W}^{(j)} \!-\! \mathbf{W}^{(j-1)}\Big|\!>\! \pmb{\varepsilon_2}$, $\mathbf{W}^{(j)}\preceq0$,  $\mathbf{W}^{(j)}\succeq1$, or $\exists_{k,n} \alpha_{k,n}^{(i)} < 0$}{step up $\eta^{(j)}$, $j\!=\!j\!+\!1$, and \textbf{goto} {\scriptsize{\textbf{1}}}.}
     Return $\pmb{\alpha}^* = \pmb{\alpha}^{(i)}$ and $\mathbf{W}^* = \mathbf{W}^{(j)}$.
\end{algorithm}

\vspace{-0.155cm}
\begin{remark}\label{remark1} The convexity of ${\mathcal{L}}$ relies on the forms of $g_1$, $g_2$, and $g_3$. For convex $g_1$ and $g_2$ but concave $g_3$, the dual problem is definitely convex with the following conditions
\begin{align}
    \mathcal{H}_4&: \dfrac{\partial \mathcal{F}_1}{\partial \alpha_{k,n}} \dfrac{\partial P_{{\rm e}, n}}{\partial \alpha_{k,n}} + \dfrac{\mathcal{F}_1}{2}  \dfrac{\partial^2 P_{{\rm e}, n}}{\partial \alpha_{k,n}^2} \geq 0, \nonumber \\
    \mathcal{H}_5&: \dfrac{\partial \mathcal{S}_n}{\partial \alpha_{k,n}} \dfrac{\partial \mathcal{F}_2}{\partial \alpha_{k,n}} + \dfrac{\mathcal{S}_n}{2}  \dfrac{\partial^2 \mathcal{F}_2}{\partial \alpha_{k,n}^2} \geq 0. \nonumber
\end{align}
For all being convex, ${\mathcal{L}}$ is convex if $\mathcal{H}_4$, $\mathcal{H}_5$, and $\mathcal{H}_6$ hold, with
\begin{align}\label{eq:eq17}
     \mathcal{H}_6 &: \eta \left(\mathcal{F}_1\dfrac{\partial \mathcal{S}_n}{\partial \alpha_{k,n}} \!+\! \mathcal{S}_n\dfrac{\partial \mathcal{F}_1}{\partial \alpha_{k,n}} \right) g_3^{\prime\prime}\!\left( \mathcal{F}_1 \mathcal{S}_n\overline{\mathcal{S}}_n\right)\!   \nonumber \\
     & ~~~~~~ \leq  w_2\left(\mathcal{F}_2\dfrac{\partial \mathcal{S}_n}{\partial \alpha_{k,n}} \!+\! \mathcal{S}_n\dfrac{\partial \mathcal{F}_2}{\partial \alpha_{k,n}} \right) g_2^{\prime\prime}\!\left( \mathcal{F}_2 \mathcal{S}_n\overline{\mathcal{S}}_n\right)\!.
\end{align}
\end{remark}
The reverse of Remark~\ref{remark1} is \emph{not} necessarily true. Exerting \eqref{eq:eq12}, \eqref{eq:eq21}, and Remark~\ref{remark1}, optimal forms of $h$ and $f_\rho$ are found.

\vspace{-0.155cm}
\subsection{Analysis for Large $I_n$}
We consider $\mathcal{X}_n$, $\forall n$, has large enough state space, i.e., $I_n \gg 1$. In this regime, we achieve $\underset{I_n \to \infty}{\lim} P_{{\rm e}, n} = \frac{1}{2}$ from \eqref{prob1:eq1}. 
Besides, we rewrite \eqref{prob1:eq2} and define ${\mathcal{E}}_{n}^{(k)}$ and $\widehat{\mathcal{E}}_{n}^{(k)}$ as below
\begin{align*}
    {\mathcal{E}}_{n}^{(k)} =  \dfrac{\sum_{\ell: k \in \mathcal{R}_\ell} \mathcal{E}_{n,\ell}}{\alpha_{k,n}}
    , ~\, \widehat{\mathcal{E}}_{n}^{(k)} =  \dfrac{\sum_{\ell: k \in \mathcal{R}^\prime_\ell} \mathcal{E}_{n,\ell}}{1 - \alpha_{k,n}}.
\end{align*}
Thereby, using \eqref{eq:eq23}, $\alpha^*_{k,n}$, $\forall k,n$, is derived as follows
\begin{align}\label{eq:largeL2}
    \alpha_{k,n}^* = \dfrac{\eta \overline{\mathcal{S}}_n {\rm{EUU}}_{\rm min}\Big(1\!- \widehat{\mathcal{E}}_{n}^{(k)}\Big) - 2w_1g_1\!\!\left( \dfrac{\mathcal{F}_1}{2^{(|\widehat{\mathcal{A}}_t|+1)}}\right)}{\eta \overline{\mathcal{S}}_n {\rm{EUU}}_{\rm min} \Big( {\mathcal{E}}_{n}^{(k)} \!- \widehat{\mathcal{E}}_{n}^{(k)}\Big)\!}.
\end{align}

From \eqref{eq:largeL2}, we infer that $\alpha_{k,n}^*$ increases by (i) decreasing the average usefulness of updates to tackle the high contamination of useful updates over the channels, (ii) enlarging the set of queried attributes, and (iii) increasing the ${\rm{EUU}}_{\rm min}$ constraint.

\vspace{-0.165cm}
\subsection{Decision Making Criterion}\label{Section4:4d}
To obtain the threshold criterion $v_{{\rm th}, k}^{(n)}$, $\forall k$, $\forall n\in\mathcal{A}_k$, for the self-decision scheme proposed in \ref{Section2:2a}, the $k$-th ISA needs to timely compute $\alpha_{k,n}^*$, $\forall k,n$, according to Section~\ref{section4:partB}. To this end, the ISA measures its distance from all NMAs thanks to the received query signals and has knowledge of $\mathcal{K}_n$, the other ISAs' location distribution, update usefulness distributions, and observation accuracy. We derive the following upper bound for $\gamma_{k,m,n}$ in \eqref{eq:eq4} by applying the Markov's inequality 
\begin{align}\label{eq:DMK1}
    &\gamma_{k, m, n}^{(\mathcal{C}, \mathcal{I})} \!\leq\!  
    \underset{\mathcal{C}_k}{\mathbb{E}} \!\left\{ \sum_{k^{\prime} \in \mathcal{C}_k} \sum_{k^{\prime\prime} \in \mathcal{I}} \!\dfrac{\prod_{k^{\prime}_0 \in \mathcal{C}_k} \Lambda_{k^{\prime}_0, m}^{(n)} \prod_{k^{\prime\prime}_0 \in \mathcal{I}} \Omega_{k^{\prime\prime}_0, m}^{(n)} }{\!\left(\Lambda_{k^{\prime}, m}^{(n)} \!+\! \Omega_{k^{\prime\prime}, m}^{(n)}\right)\!\!\left(\Lambda_{k^{\prime}, m}^{(n)}\right)^{\!3}}\,\Omega_{k^{\prime\prime}, m}^{(n)} \!\right\}\!.
\end{align}

After equalizing $\gamma_{k, m, n}^{(\mathcal{C}, \mathcal{I})}$ to its upper bound and using Algorithm~\ref{Alg:Alg.1}, $\alpha_{k,n}^*$ is derived for the $k$-th ISA and $n$-th attribute at any arbitrary service interval. Eventually, we arrive 
\begin{align}
    v_{{\rm th}, k}^{(n)} = F_{V}^{-1}\!\!\left(1\!-\! \operatorname{min}\!\left\{1,\dfrac{\alpha_{k,n}^*}{\operatorname{max}\!\big\{0, \beta_{k,n}\big\}}\right\} \right)
\end{align}
where $F_{V}(v_{k,j}^{(n)})$ indicates the CDF of the update's meta value. Also, $\beta_{k,n}$ shows the scheme-dependent probability that the $k$-th ISA generates updates for the $n$-th attribute, given by
\begin{align}
    \beta_{k,n} = 
    \begin{cases}
    1; \hspace{41mm} \text{uniform} \\
    (I_n\!-\!1)p_n;\hspace{28mm} \text{change-aware}\\
    P_{{\rm e} , n}(1\!-\!I_np_n)+ (I_n\!-\!1)p_n; \hspace{2.6mm}
    \text{semantics-aware}
    \end{cases}
\end{align}
if $k \in \mathcal{K}_n$; otherwise, $\beta_{k,n}=0$.

\section{Simulation Results}
We investigate the performance of the proposed self-decision scheme for satisfying the goal with ${\rm{EUU}}_{\rm min}=0.1$. We consider $K=10$, $M=4$, $I_n=10$, $p^\prime_n=0.2$, $\forall n$, $|\widehat{\mathcal{A}}_t|=10$, $M_t = \lceil \frac{M+1}{2} \rceil=3$, $\forall t$, and $q_{k,n}=0.8$, $\forall k,n$. We also randomly select a subset of the ISAs constructing $\mathcal{K}_n$, $\forall n$. Additionally, we assume that the distance between each pair of ISA and NMA having a $7\,[\text{m}]$ height follows a standard normal distribution with standard deviation $60\,[\text{m}]$, the path-loss exponent $a=3.8$, $\sigma_m^2=-120\,[\text{dBm}]$, and $\gamma_{\rm th} = 10\,[\text{dB}]$. The assignment of a meta value to each update is based on a beta distribution with both shape parameters equal to $2$. Moreover, $g_1$, $g_2$, and $g_3$ have exponential forms, while $h$ and $f_\rho$ are linear and cubic functions, respectively. To simplify, we assume $\alpha_{k,n}=\alpha_{k^\prime,n^\prime}$ and accordingly $v_{{\rm th}, k}^{(n)}=v_{{\rm th}, k^\prime}^{(n^\prime)}$, $\forall n, n^\prime,k, k^\prime$.

\begin{figure}
    \centering
    \includegraphics[width=0.38\textwidth]{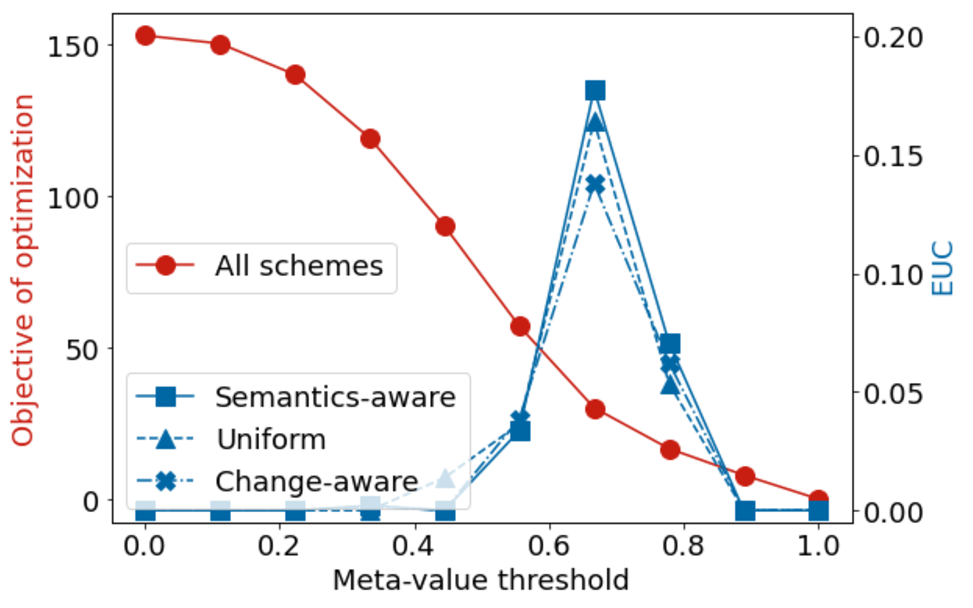}
    \vspace{-0.11cm}
    \caption{The objective of $\mathcal{P}_2$ and its constraint versus the meta value threshold.}
    \label{Fig:Figure1}
\end{figure}

Fig.~\ref{Fig:Figure1} illustrates the effect of the meta value threshold, i.e., $v_{{\rm th}, k}^{(n)}$, $\forall n,k$, hence activation probability, on the effectiveness. The optimization objective of the effectiveness problem, i.e., $\mathcal{G}$ in $\mathcal{P}_2$, and its EUU constraint are plotted versus $v_{{\rm th}, k}^{(n)}$ based on different update acquisition schemes and $\mathbf{W}=[0.5, 0.5]^T$ using Monte Carlo simulations with $500$ iterations. Despite almost identical $\mathcal{G}$, the semantics-aware scheme offers the highest EUU on average. To satisfy ${\rm{EUU}}_{\rm min}$, as shown in Fig.~\ref{Fig:Figure1}, the semantics-aware, uniform, and change-aware schemes require transmission rates of $\alpha_{k,n}=0.6$, $0.61$, and $0.62$, $\forall k,n$, respectively, under the first root, while they generate updates with rates of $0.91$, $1$, and $0.8$. Thus, we infer that the semantics-aware scheme poses, averagely, $6.4\%$ less channel load than the uniform scheme and $5\%$ more than the change-aware one.

To assess the self-decision scheme, we apply Algorithm~\ref{Alg:Alg.1} to derive the set of optimal activation probabilities based on different schemes and the assumed $\mathbf{W}$. In this case, we reach $\alpha_{k,n}=0.64$, $0.64$, and $0.67$, $\forall n,k$, for the semantics-aware, uniform, and change-aware schemes, respectively. Compared to the derived values from Fig.~\ref{Fig:Figure1}, the accuracy of the self-decision scheme is more than $92\%$. More importantly, ${\rm{EUU}}_{\rm min}$ is not satisfied without our self-decision scheme.

\begin{figure}
    \centering
    \includegraphics[width=0.335\textwidth]{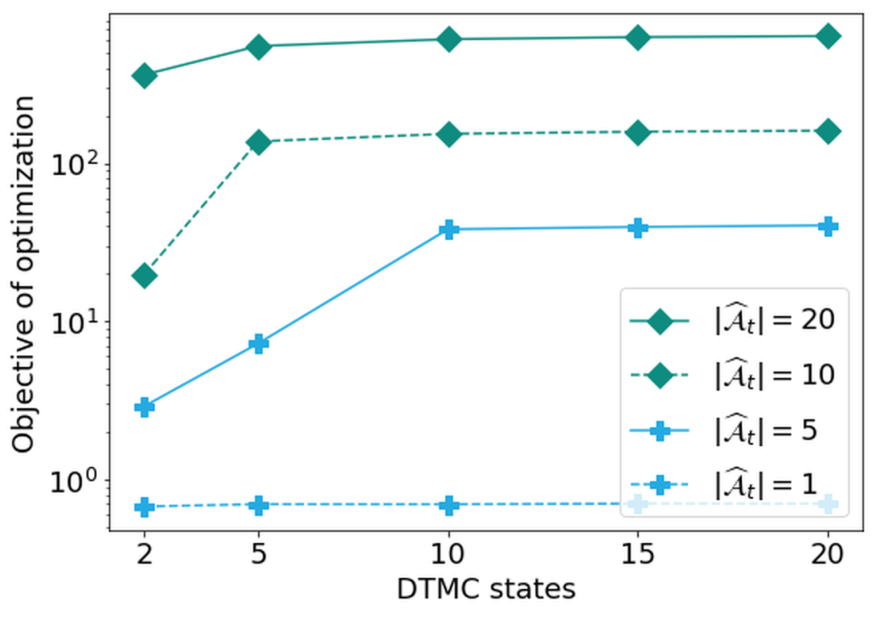}
    \vspace{-0.11cm}
    \caption{The interplay between the objective of $\mathcal{P}_2$ and the number of states in the DTMC for different sizes of required attributes.}
    \label{Fig:Figure2}
\end{figure}

The interplay between $\mathcal{G}$ and the number of states in the DTMC, i.e., $I_n$, $\forall n$, is depicted in Fig.~\ref{Fig:Figure2} based on different values of $|\widehat{\mathcal{A}}_t|$, $\forall t$. Therein, we plot the lowest value of the objective that meets ${\rm{EUU}}_{\rm min}$. We see that increasing $I_n$ results in higher $\mathcal{G}$ due to the increase of $P_{{\rm e}, n}$ and EDE, accordingly, concerning \eqref{prob1:eq1}. Additionally, higher values of $|\widehat{\mathcal{A}}_t|$ boost ERC according to \eqref{eq:eq12}, leading to higher $\mathcal{G}$. Thus, the fewer attributes queried within a service interval, the higher the effectiveness.

\section{Conclusion}
We developed a self-decision goal-oriented multiple access scheme for networked intelligent systems. Aiming to maximize the GoE, we obtained the optimal activation probabilities and threshold criteria for decision-making under different update acquisition schemes. Analyzing how different parameters affect the effectiveness, our simulation results showed that our proposal could achieve at least $92\%$ of the optimal performance.

\bibliographystyle{IEEEtran}
\bibliography{References.bib}
\balance

\end{document}